\documentclass[sigconf]{acmart}
\usepackage{booktabs}
\usepackage{enumitem}

\newcommand{\ignore}[1]{}
\newcommand{\an}[1]{$^#1$}

\copyrightyear{2019}
\acmYear{2019}
\setcopyright{acmlicensed}
\acmConference[SIGIR '19] {42nd International ACM SIGIR Conference on Research and Development in Information Retrieval}{July 21--25, 2019}{Paris, France}
\acmBooktitle{42nd Int'l ACM SIGIR Conference on Research and Development in Information Retrieval (SIGIR '19), July 21--25, 2019, Paris, France}
\acmPrice{15.00}
\acmDOI{}
\acmISBN{}

\begin{document}

\title[The Impact of Score Ties on Repeatability in Document Ranking]{The Impact of Score Ties on Repeatability\\ in Document Ranking}

\author{Jimmy Lin\an{1} and Peilin Yang}
\affiliation{\vspace{0.1cm}
  \institution{$^{1}$ David R. Cheriton School of Computer Science, University of Waterloo}
}
\email{jimmylin@uwaterloo.ca}

\renewcommand{\shortauthors}{Jimmy Lin and Peilin Yang}

\begin{abstract}
Document ranking experiments should be repeatable.
However, the interaction between multi-threaded indexing and score ties during retrieval may yield non-deterministic rankings, making repeatability not as trivial as one might imagine.
In the context of the open-source Lucene search engine, score ties are broken by internal document ids, which are assigned at index time.
Due to multi-threaded indexing, which makes experimentation with large modern document collections practical, internal document ids are not assigned consistently between different index instances of the same collection, and thus score ties are broken unpredictably.
This short paper examines the effectiveness impact of such score ties, quantifying the variability that can be attributed to this phenomenon.
The obvious solution to this non-de\-ter\-min\-ism and to ensure repeatable document ranking is to break score ties using external collection document ids.
This approach, however, comes with measurable efficiency costs due to the necessity of consulting external identifiers during query evaluation.
\end{abstract}

\maketitle
\fancyhead{}

\section{Introduction}

It should generate no controversy to assert that repeatability of document ranking experiments in information retrieval research is a desirable property.
To be precise, running the same ranking model over the same document collection with the same queries should yield the same output every time.
Yet, this simple property is not trivial to achieve in modern retrieval engines that take advantage of multi-threaded indexing.
In this paper, we explore corner cases that yield non-repeatable rankings:\
observed non-determinism is attributable to score ties, or documents in the collection that receive the same score with respect to a particular ranking model.

Anserini, an open-source information retrieval toolkit built on Lucene~\cite{Yang_etal_SIGIR2017,Yang_etal_JDIQ2018}, provides the context for our study.
The system evolved from previous IR reproducibility experiments~\cite{Arguello_etal_SIGIRForum2015,Lin_etal_ECIR2016} where Lucene exhibited a good balance between efficiency and effectiveness compared to other open-source search engines.
A large user and developer base, numerous commercial deployments at scale, as well as a vibrant ecosystem provide additional compelling arguments for building an IR research toolkit on top of Lucene.

The multi-threaded indexer that Anserini implements on top of Lucene is able to rapidly index large modern document collections---for example, building a simple non-positional index on the popular ClueWeb small collections each takes around an hour on a typical server~\cite{Yang_etal_JDIQ2018}.
A consequence of the multi-threaded design is that documents are inserted into the index in a non-deterministic order, which means that different index instances over the same collection may be substantively different.
This has implications for documents that receive the same score at retrieval time---by default, the Lucene query evaluation algorithm breaks ties by an internal document id, which is assigned based on document insertion order.
Since these internal document ids are not stable across different index instances, ranking experiments may not be repeatable.

Is this a big deal? We argue {\it yes}, from a number of perspectives.
While arbitrary tie-breaking behavior has a relatively small impact on simple ``bag of words'' queries (typically, differences in the fourth decimal place in terms of standard evaluation metrics), effectiveness differences can be magnified for relevance feedback runs that utilize two-stage retrieval based on an initial ranking.
Repeatable runs form a cornerstone of regression testing in modern software development---without exact repeatability, it is difficult to separate non-deterministic bugs (so called ``heisenbugs'') from inherent execution instability.
Without a suite of regression tests, sustained progress on a complex codebase becomes difficult.
For example, Lin et al.~\cite{Lin_etal_ECIR2016} reported cases of different results from runs that purport to use the same ranking model from the same system on the same test collection (by the same research group, even).

The goal of this paper and our contribution is a detailed study of the impact of score ties from the perspective of repeatability across a number of different test collections for a specific search engine.
We empirically characterize differences in effectiveness that can be attributed to arbitrary interleaving of documents ingested during multi-threaded indexing.
The solution to repeatable document ranking is fairly obvious:\ ties should be broken deterministically by external collection document ids (which are stable) instead of internal indexer-assigned ids.
However, this comes at a measurable cost in query evaluation performance, which arises from the need to consult external identifiers during query evaluation.

\section{Experimental Design}

All experiments in this paper were conducted with Anserini v0.1.0, which is based on Lucene 6.3.0; all code necessary to replicate our experiments are available on GitHub at {\tt http://anserini.io/}.

To examine the impact of score ties across a diverse range of document types, we considered three newswire collections, two tweet collections, and two web collections:

\vspace{-0.02cm}
\begin{itemize}[leftmargin=*]

\item TREC 2004 Robust Track, on TREC Disks 4 \& 5.

\item TREC 2005 Robust Track, on the AQUAINT Corpus.

\item TREC 2017 Common Core Track, on the New York Times Annotated Corpus.

\item TREC 2011/2012 Microblog Tracks (on the Tweets2011 collection)
  and TREC 2013/2014 Microblog Tracks (on the Tweets2013 collection).

\item TREC 2010--2012 Web Tracks (on ClueWeb09b) and
  TREC 2013--2014 Web Tracks (on ClueWeb12-B13).

\end{itemize}

\begin{table*}[t]
\centering TREC 2004 Robust Track topics, Disks 4 \& 5\\[0.5ex]
\centering
\begin{tabular}{|l|rrr|rrr|}
\hline
Model & AP & min---max & $\Delta$ & P30 & min---max & $\Delta$ \\
\hline
\hline
BM25     & 0.2501 & 0.2498 --- 0.2501 & 0.0003 & 0.3123 & 0.3120 --- 0.3124 & 0.0004 \\
BM25+RM3 & 0.2757 & 0.2756 --- 0.2757 & 0.0001 & 0.3256 & 0.3253 --- 0.3257 & 0.0004 \\
QL       & 0.2468 & 0.2464 --- 0.2469 & 0.0005 & 0.3083 & 0.3076 --- 0.3080 & 0.0007 \\
QL+RM3   & 0.2645 & 0.2643 --- 0.2644 & 0.0002 & 0.3153 & 0.3149 --- 0.3151 & 0.0004 \\
\hline
\end{tabular}

\bigskip

\centering TREC 2005 Robust Track topics, AQUAINT Collection\\[0.5ex]
\centering
\begin{tabular}{|l|rrr|rrr|}
\hline
Model & AP & min---max & $\Delta$ & P30 & min---max & $\Delta$ \\
\hline
\hline
BM25     & 0.2003 & 0.2000 --- 0.2006 & 0.0006 & 0.3660 & 0.3660 --- 0.3673 & 0.0013 \\
BM25+RM3 & 0.2511 & 0.2506 --- 0.2513 & 0.0007 & 0.3873 & 0.3860 --- 0.3880 & 0.0020 \\
QL       & 0.2026 & 0.2019 --- 0.2026 & 0.0005 & 0.3713 & 0.3693 --- 0.3720 & 0.0027 \\
QL+RM3   & 0.2480 & 0.2471 --- 0.2483 & 0.0012 & 0.4007 & 0.4007 --- 0.4013 & 0.0006 \\
\hline
\end{tabular}

\bigskip

\centering TREC 2017 Common Core Track topics, New York Times Collection\\[0.5ex]
\centering
\begin{tabular}{|l|rrr|rrr|}
\hline
Model & AP & min---max & $\Delta$ & P30 & min---max & $\Delta$ \\
\hline
\hline
BM25     & 0.1996 & 0.1997 --- 0.1998 & 0.0002 & 0.4207 & 0.4213 --- 0.4220 & 0.0007 \\
BM25+RM3 & 0.2633 & 0.2632 --- 0.2635 & 0.0003 & 0.4893 & 0.4867 --- 0.4893 & 0.0026 \\
QL       & 0.1928 & 0.1929 --- 0.1929 & 0.0001 & 0.4327 & 0.4327 --- 0.4333 & 0.0006 \\
QL+RM3   & 0.2409 & 0.2408 --- 0.2409 & 0.0001 & 0.4647 & 0.4640 --- 0.4647 & 0.0007 \\
\hline
\end{tabular}

\vspace{0.15cm}
\caption{Variability in effectiveness attributed to score ties on three newswire collections.
The first column for each metric (``AP'' and ``P30'') shows values with consistent tie-breaking.
Columns marked ``min---max'' report minimum and maximum scores across five different indexes with arbitrary tie-breaking.
Columns marked ``$\Delta$'' report the largest absolute observed difference (including consistent tie-breaking).}
\label{results:newswire}
\vspace{-0.3cm}
\end{table*}

\begin{table*}[t]
\centering TREC 2011 and 2012 Microblog Track topics, Tweets2011 Collection\\[0.5ex]
\centering
\begin{tabular}{|l|rrr|rrr|}
\hline
Model & AP & min---max & $\Delta$ & P30 & min---max & $\Delta$ \\
\hline
\hline
QL       & 0.2787 & 0.2761 --- 0.2770 & 0.0026 & 0.3673 & 0.3636 --- 0.3667 & 0.0037 \\
QL+RM3   & 0.3178 & 0.3157 --- 0.3173 & 0.0021 & 0.3954 & 0.3929 --- 0.3975 & 0.0046 \\
\hline
\end{tabular}

\bigskip

\centering TREC 2013 and 2014 Microblog Track topics, Tweets2013 Collection\\[0.5ex]
\centering
\begin{tabular}{|l|rrr|rrr|}
\hline
Model & AP & min---max & $\Delta$ & P30 & min---max & $\Delta$ \\
\hline
\hline
QL       & 0.3357 & 0.3345 --- 0.3348 & 0.0012 & 0.5429 & 0.5406 --- 0.5423 & 0.0023 \\
QL+RM3   & 0.3692 & 0.3695 --- 0.3699 & 0.0007 & 0.5484 & 0.5528 --- 0.5542 & 0.0058 \\
\hline
\end{tabular}

\vspace{0.15cm}
\caption{Variability in effectiveness attributed to score ties on tweet collections,
organized in the same way as Table~\ref{results:newswire}.}
\label{results:tweets}
\vspace{-0.4cm}
\end{table*}

\noindent For each document collection, we used Anserini to build five separate indexes from scratch.
For each index, we performed a retrieval run using topics from the corresponding test collections.
In each of these runs, Anserini used Lucene's default tie-breaking technique based on arbitrarily-assigned internal document ids---which as we have noted above, is not consistent between index instances due to multi-threading.
Differences in effectiveness between these runs quantify the impact of score ties.

In Anserini, we have modified the query evaluation algorithm to use the external collection id to break score ties, which means that retrieval results are repeatable across different index instances.
This is accomplished via the \texttt{Sort} abstraction in Lucene, which allows the developer to specify how ranked lists are sorted.
Naturally, the default is by the score produced by the ranking model.
For newswire and web collections, we added lexicographic sort order of collection document ids as the tie breaker.
For tweets, ties are broken by reverse chronological order (i.e., most recent tweet first).
In our experiments, we refer to this as the {\it repeatable} ranking condition, which provides the ``ground truth'' for comparison against Lucene's default tie-breaking behavior above.

With the exception of tweet collections, we considered the following ranking models:\ BM25 and query likelihood, and the RM3 query expansion technique applied to both.
For tweet collections, we only considered query likelihood since BM25 is known not to be effective.
In Anserini, RM3 is implemented as a two-stage process:\ a relevance model is estimated from documents in an initial ranked list, which then forms an expanded query that retrieves the final results.
Thus, there are two source of variability due to score ties---when applying a rank cutoff in the initial retrieval as well as the final ranking.

All runs retrieved up to 1000 hits and were evaluated in terms of standard retrieval metrics:\
for newswire and tweet collections, we computed average precision (AP) and precision at rank 30 (P30) using {\tt trec\_eval}.
For the web collections, we computed NDCG@20 using {\tt gdeval.pl} (since the shallow pool depths make AP unreliable).
As a final detail, to ensure that the evaluation program respects our tie-breaking approach, Anserini slightly perturbs scores of tied documents when writing output in TREC format so that the rank order is consistent with the score order.
This is necessary, for example, because when {\tt trec\_eval} computes a metric, it loads in the run and sorts the documents by score, breaking ties internally.
We did not want an external evaluation tool to impose its own tie-breaking approach, but rather respect the rankings generated by our system.

\begin{table}[t]
\vspace{0.12cm}
\centering TREC 2010---2012 Web Track topics, ClueWeb09b \\[0.5ex]
\centering
\begin{tabular}{|l|rrr|}
\hline
Model & NDCG@20 & min---max & $\Delta$ \\
\hline
\hline
BM25     & 0.1407 & 0.1405 --- 0.1408 & 0.0003 \\
BM25+RM3 & 0.1524 & 0.1524 --- 0.1525 & 0.0001 \\
QL       & 0.1211 & 0.1210 --- 0.1212 & 0.0002 \\
QL+RM3   & 0.1340 & 0.1340 --- 0.1342 & 0.0002 \\
\hline
\end{tabular}

\bigskip

\centering TREC 2013 and 2014 Web Track topics, ClueWeb12-B13 \\[0.5ex]
\centering
\begin{tabular}{|l|rrr|}
\hline
Model & NDCG@20 & min---max & $\Delta$ \\
\hline
\hline
BM25     & 0.1216 & 0.1216 --- 0.1216 & 0.0000 \\
BM25+RM3 & 0.1080 & 0.1077 --- 0.1083 & 0.0006 \\
QL       & 0.1146 & 0.1146 --- 0.1154 & 0.0008 \\
QL+RM3   & 0.0920 & 0.0920 --- 0.0926 & 0.0006 \\
\hline
\end{tabular}

\vspace{0.15cm}
\caption{Variability in effectiveness attributed to score ties on web collections,
organized in the same way as Table~\ref{results:newswire}.}
\label{results:web}
\vspace{-0.5cm}
\end{table}

\section{Results}

The results of our experiments on the newswire collections are shown in Table~\ref{results:newswire}.
Under the columns with the names of the metric (``AP'' and ``P30''), we report the effectiveness of the repeatable runs (consistent tie-breaking with external document ids).
The columns marked ``min---max'' report minimum and maximum scores across the five different indexes given arbitrary tie-breaking (internal document ids).
Note that for some cases, the effectiveness of the repeatable condition falls outside the min--max range.
The columns marked ``$\Delta$'' show the largest absolute observed difference across all runs, including the repeatable condition.
Results on the tweet collections are shown in Table~\ref{results:tweets} and results on the web collections are shown in Table~\ref{results:web};
both are organized in exactly the same manner as Table~\ref{results:newswire}.
Note that for the web collections we only report effectiveness in terms of NDCG@20 due to the shallow pools used in the construction of the qrels.

We see that the variability attributed to tie-breaking behavior yields minor effectiveness differences, usually in the fourth decimal place, but sometimes in the third decimal place.
Overall, observed variability in AP is smaller than P30 because for AP, the differences come from documents that straddle the rank cutoff of 1000, where score contributions are small.
Results show that RM3 can exhibit (but not always) greater variability because score ties impact both the selection of documents for extracting feedback terms as well as the final ranking.
For tweets, we observe greater variability, due to more score ties since many tweets have the same length.
Differences in P30 are more pronounced than AP.
Results on the web collections are consistent with the other collections.

In absolute terms, the observed score variability is fairly small.
However, to put these differences in perspective, incremental advances in many popular NLP and IR shared tasks today, for example, SQuAD and MS MARCO, are measured in the third decimal place.
Leaving aside whether such leaderboard-driven research is good for the community as a whole, we note that the amount of variability observed in our experiments can approach the magnitude of differences in successive advances in ``the state of the art''.

More importantly, as argued in the introduction, this variability makes regression testing---which is a cornerstone of modern software development---very difficult.
Typically, in regression testing, for floating point values the developer specifies a tolerance when comparing test results with expected results, for example, to compensate for precision errors.
In our case, it is not clear how the developer should set this tolerance.
A value too large would fail to catch genuine bugs, while a value too small would cause frequent needless failures, defeating the point of regression testing.

As discussed in the previous section, the solution to repeatable document ranking is relatively straightforward---instead of depending on the internal document id to break score ties, we should use external (collection-specific) document ids.
Anserini implements exactly this solution using a Lucene-provided abstraction, as described in the previous section.
This approach, however, comes at a cost in terms of efficiency, since the query evaluation algorithm must consult an external id as part of its inner loop during postings traversal.
Lookup of an external id requires some form of memory access, likely leading to pointer chasing that can potentially disrupt data and cache locality.

The efficiency costs of repeatable experiments are quantified in Table~\ref{results:latency} for the three largest collections used in our experiments.
Here, we report average query evaluation latency (in seconds) under the non-repeatable and repeatable conditions, averaged over five trials on an iMac Pro desktop machine (2.3 GHz Intel Xeon W processor) running macOS High Sierra.
The final column shows the increase in query latency due to consistent tie-breaking using external document ids.
In all cases we first ran the experiments a few times to warm up underlying operating system caches, and then captured measurements over the next sets of trials.
Query evaluation was performed using a single thread.

For simple bag-of-words queries, we observe a measurable slowdown in query latency, which quantifies the cost of repeatability.
Across the web collections, this slowdown is approximately 20\%, but for tweets the latency costs are a bit higher, most likely due to more prevalent score ties.
Not surprisingly, query evaluation with RM3 is much slower due to its two-stage process:\ here, however, the behavior between tweet and web collections diverge.
For web collections, the slowdown is less compared to bag-of-words queries because a significant amount of time is spent reading document vectors from the index and estimating relevance models during query evaluation.
As a result, the amount of time actually spent in the postings traversal inner loop is proportionally smaller.
Tweets, however, are much shorter, and so estimating relevance models is relatively fast.
The larger expanded queries require consulting more postings and scoring more documents, thus leading to large slowdowns for repeatable runs.

\begin{table}[t]
\centering TREC 2013 and 2014 Microblog Track topics, Tweets2013 \\[0.5ex]
\centering
\begin{tabular}{|l|rr|r|}
\hline
Model & Non-Repeatable & Repeatable & $\Delta$ \\
\hline
\hline
QL     & 0.46s & 0.58s & $+$26\%\\
QL+RM3 & 1.05s & 2.00s & $+$90\% \\
\hline
\end{tabular}

\vspace{0.25cm}

\centering TREC 2010---2012 Web Track topics, ClueWeb09b \\[0.5ex]
\centering
\begin{tabular}{|l|rr|r|}
\hline
Model & Non-Repeatable & Repeatable & $\Delta$ \\
\hline
\hline
BM25     & 0.18s & 0.23s & $+$23\%\\
BM25+RM3 & 3.92s & 4.18s & $+$7\% \\
\hline
\end{tabular}

\vspace{0.25cm}

\centering TREC 2013---2014 Web Track topics, ClueWeb12-B13 \\[0.5ex]
\centering
\begin{tabular}{|l|rr|r|}
\hline
Model & Non-Repeatable & Repeatable & $\Delta$ \\
\hline
\hline
BM25     & 0.22s & 0.26s & $+$18\%\\
BM25+RM3 & 4.23s & 4.61s & $+$9\% \\
\hline
\end{tabular}

\vspace{0.15cm}

\caption{Latency differences between non-repeatable and repeatable
  document ranking, where repeatability is achieved by consistently
  breaking ties using external document ids.}
\label{results:latency}
\vspace{-0.6cm}
\end{table}

We acknowledge that these results are implementation specific, tied to the exact mechanism by which external document ids are consulted.
Our current implementation uses existing Lucene abstractions for controlling the sort order of results, 
but greater efficiencies might be possible with a more invasive modification of Lucene internals.
Nevertheless, our broader point remains true:\ repeatability inevitably comes at some cost in performance.

\section{Related Work}

We are, of course, not the first to have noticed score ties in document ranking and to examine their impact.
Cabanac et al.~\cite{Cabanac_etal_CLEF2010} studied the behavior of the widely-used \texttt{trec\_eval} tool (also used in our study) and concluded that tie-breaking strategy can have substantive impact on conclusions about the relative effectiveness of different runs in terms of average precision.
More recently, Ferro and Silvello~\cite{Ferro_Silvello_etal_ECIR2015} found similar issues in their study of rank-biased precision.
McSherry and Najork~\cite{McSherry_Najork_ECIR2008} proposed efficient ways to compute various IR metrics in the presence of score ties (predating both above papers); broader adoption would have helped address many of the issues in this paper, but these techniques have not gained traction.
The organizers of the TREC Microblog Tracks~\cite{Ounis_etal_TREC2011} have discovered that tie-breaking heuristics can have a large impact on effectiveness, since tweets are short and hence many retrieved results share the same score.
In particular, reverse chronological sorting of tweets with the same score in ranked retrieval increases effectiveness and makes sense from the task perspective.

While most of the above cited papers focus on the implications of scoring ties for IR evaluation, others have examined different aspects of the phenomenon.
For example, Wu and Fang~\cite{Wu_Fang_ICTIR2013} used score ties to prioritize relevance signals in document ranking.
Z.~Yang et al.~\cite{Yang_etal_AIR2016} explored different levels of score rounding as a way to accelerate query processing---for example, taking advantage of approximate scoring regimes.
This relates to impact quantization~\cite{Anh_etal_SIGIR2001}, and JASS~\cite{Lin_Trotman_ICTIR2015} is an example of a recent system that exploits approximate scoring for anytime ranking.

This study builds on previous work, but examines a new angle that to our knowledge has not been explored---the impact of score ties from the perspective of experimental repeatability.
Recent ACM guidelines\footnote{https://www.acm.org/publications/policies/artifact-review-badging} articulate the role of repeatability as an important foundation of scientific methodology with computational artifacts.
Without repeatability, replicability and reproducibility are not possible.
Thus, our work makes a contribution towards firmly establishing repeatability in IR experiments using Lucene.

\section{Conclusions}

The conclusions from our examination of score ties are fairly clear:\
Although absolute differences in effectiveness metrics are relatively small---in the third decimal place at the most---these differences nevertheless pose challenges for regression testing.
Without rigorous regression testing, it is difficult to put progress on solid footing in terms of software engineering best practices, since developers cannot be certain if a new feature introduced a bug.
Fortunately, the solution to repeatable runs is fairly straightforward, which we have implemented:\ score ties should be broken by external collection ids.
However, this comes with a measurable efficiency cost in terms of slowdown in query evaluation.
As a concrete recommendation for navigating this tradeoff, we suggest that non-repeatable runs are acceptable for prototyping, but any permanent contributions to a codebase must pass slower regression tests that make repeatability a requirement.

\smallskip \noindent {\bf Acknowledgments.} This work was
supported in part by the Natural Sciences and Engineering Research
Council (NSERC) of Canada.


\end{document}